\documentclass[10pt,a4paper,twocolumn,english,superscriptaddress,nobalancelastpage,prl]{revtex4}
\usepackage[latin9]{inputenc}
\usepackage{textcomp}
\usepackage{amsmath}
\usepackage{amssymb}
\usepackage{graphicx}

\makeatletter

\pdfpageheight\paperheight
\pdfpagewidth\paperwidth

\@ifundefined{textcolor}{}
{%
 \definecolor{BLACK}{gray}{0}
 \definecolor{WHITE}{gray}{1}
 \definecolor{RED}{rgb}{1,0,0}
 \definecolor{GREEN}{rgb}{0,1,0}
 \definecolor{BLUE}{rgb}{0,0,1}
 \definecolor{CYAN}{cmyk}{1,0,0,0}
 \definecolor{MAGENTA}{cmyk}{0,1,0,0}
 \definecolor{YELLOW}{cmyk}{0,0,1,0}
}

\usepackage{amsfonts}\usepackage{eurosym}\usepackage{amsthm}

\makeatother

\usepackage{babel}
\begin{document}

\title{Electric quantum walks in two dimensions}

\author{Luis A. Bru}

\affiliation{Optical and Quantum Communications Group, ITEAM Research Institute,
Universitat Politècnica de València, Camino de Vera s/n, 46022--València,
Spain}

\author{Margarida Hinarejos}

\affiliation{Instituto de Física, Facultad de Ingeniería, Universidad de la República,
CC 30, CP 11000, Montevideo, Uruguay}

\author{Fernando Silva, Germán J. de Valcárcel, and Eugenio Roldán}

\affiliation{Departament d'Òptica, Universitat de València, Dr. Moliner 50, 46100--Burjassot,
Spain}
\begin{abstract}
We study electric quantum walks in two dimensions considering Grover,
Alternate, Hadamard, and DFT quantum walks. In the Grover walk the
behaviour under an electric field is easy to summarize: when the field
direction coincides with the $x$ or $y$ axes, it produces a transient
trapping of the probability distribution along the direction of the
field, while when it is directed along the diagonals, a perfect 2D
trapping is frustrated. The analysis of the alternate walk helps to
understand the behaviour of the Grover walk as both walks are partially
equivalent; in particular, it helps to understand the role played
by the existence of conical intersections in the dispersion relations,
as we show that when these are removed a perfect 2D trapping can occur
for suitable directions of the field. We complete our study with the
electric DFT and Hadamard walks in 2D, showing that the latter can
exhibit perfect 2D trapping. 
\end{abstract}
\maketitle

\section{Introduction}

The quantum walk (QW) \cite{Kempe03,Kendon06,Konno08,Venegas12} is
a simple quantum diffusion model that can be understood as the quantum
analog of classical random walks \cite{Strauch06}, but also as the
discretized version of different wave equations including the Schrödinger
\cite{deValcarcel10,Hinarejos13} and the Dirac \cite{Strauch07,Chandrasekar13,DiMolfetta14}
equations. Appearing in two basic forms, namely the coined QW \cite{Aharonov93,Meyer96}
and the continuous QW \cite{Farhi98,Childs04}, here we deal with
the former.

Originally introduced by Feynman \cite{Feynman65}, QWs were rediscovered
many years later by different authors in different contexts \cite{Aharonov93,Meyer96,Watrous01,Farhi98,Childs04},
and maybe not surprisingly they are finding very diverse applications
and unsuspected connections, nonetheless the fact that QWs have been
shown to constitute a universal model for quantum computation in both
its continuous \cite{Childs09} and discrete \cite{Lovett10} forms.
Along the last decade or so there have been a number of experimental
researches in quite diverse platforms (see \cite{review exp} for
a recent review), including ``classical'' devices \cite{Bouwmeester99},
as QWs can be implemented by using classical means only, at least
for a not too large dimensionality of the QW \cite{Knight03,Roldan05}.

One interesting aspect to consider in QWs is the effect of external,
artificial fields such as gravitational, magnetic, or electric fields.
While gravitational \cite{DiMolfetta14} or magnetic \cite{magnetic QW,magnetic2}
QWs have been considered only recently, electric QWs (EQW) already
were considered by Meyer\cite{Meyer97,Meyer98} and have been the
subject of continued attention at least for a decade, and are by now
quite well understood in the one dimensional (1D) case \cite{Wojcik04,Banuls06,Regensburger11,Cedzich13,Genske13,Xue15}.
We mention also related work on other inhomogeneous QWs \cite{Linden09,Shikano10}.

In the 1D-EQW the quantum walker acquires an additional phase $\varphi$
(proportional to the applied electric field amplitude) that depends
on the walker position. It turns out that the behaviour of the 1D--EQW
is very sensitive to the value of the additional phase. While, in
general, a non null $\varphi$ leads (roughly) to localization phenomena,
the detailed behaviour, however, depends on whether $\varphi/2\pi$
is a rational or an irrational number, which somehow connects the
1D-EQW with 2D conductors under the action of a normal magnetic field,
where similar dependences on the rationality, or not, of a parameter
manifest in the celebrated Hofstadter butterfly \cite{Hofstadter}.
In this article we go beyond the 1D case and study 2D electric QWs
(2D-EQWs) in some detail, a problem previously considered in \cite{DiFranco15}
from a different perspective.

The rest of the article organizes as follows: in Section II we review
what is known for 1D-EQWs, and then in Section III we study 2D-EQWs.\ We
first consider the Grover walk, then the alternate 2D-EQW (for which
some analytical results can be derived), and in a last subsection
those using the DFT and the Hadamard coins. In particular, we can
understand what is necessary for a perfect 2D-trapping of the probability
distribution: the absence of conical intersections in the walk dispersion
relations. Finally, in Section IV, we give our main conclusions.

\section{Electric QWs in 1D}

Consider a walker with an attached qubit, the coin, moving along the
one--dimensional infinite lattice. The state at\ time $t$ can be
written as $\left\vert \psi\left(t\right)\right\rangle =\sum_{x}\sum_{s}a_{x,s}\left(t\right)\left\vert x,s\right\rangle $
where $x\in Z$ denotes position and $s=\pm1$ denotes the coin state,
so that $\left\vert a_{x,s}\left(t\right)\right\vert ^{2}$ is the
probability of finding the walker at (discrete) time $t$ at (discrete)
position $x$ with coin state $s$. Of course $\sum_{x}\sum_{s}\left\vert a_{x,s}\left(t\right)\right\vert ^{2}=1$
for all $t$.

The state evolves according to $\left\vert \psi\left(t+1\right)\right\rangle =\hat{U}_{\varphi}\left\vert \psi\left(t\right)\right\rangle $
with $\hat{U}_{\varphi}$ the unitary evolution operator
\begin{equation}
\hat{U}_{\varphi}\equiv e^{i\varphi\hat{x}}\hat{U}_{0}=e^{i\varphi\hat{x}}\hat{S}\hat{C},\label{Ufi}
\end{equation}
being $\hat{U}_{0}$ the standard QW evolution operator and $e^{i\varphi\hat{x}}$
the electric field unitary. The conditional shift operator reads
\begin{eqnarray}
\hat{S} & = & \sum_{x=-\infty}^{+\infty}\sum_{s=\pm1}\left\vert x+s,s\right\rangle \left\langle x,s\right\vert ,\label{S}
\end{eqnarray}
the coin operator acts onto the qubit parts of the state and can be
represented by the following matrix

\begin{equation}
\hat{C}=\begin{pmatrix}e^{i\left(\alpha+\beta\right)}\cos\theta & e^{i\left(\alpha-\beta\right)}\sin\theta\\
-e^{-i\left(\alpha-\beta\right)}\sin\theta & e^{-i\left(\alpha+\beta\right)}\cos\theta
\end{pmatrix},\label{eq:C}
\end{equation}
wich can be understood as a general unitary rotation parametrized
by the three angles $\alpha$,$\,\beta\text{, and }\theta$. It turns
out that the unique dynamically relevant parameter is $\theta$ \cite{Roldan13},
and setting the phases $\alpha$ and $\beta$ to zero just fixes the
rotation axis on the Bloch sphere to be the $y$-axis, i.e. $\hat{C}=\exp\left\{ i\theta\hat{\sigma}_{2}\right\} $
with $\hat{\sigma}_{2}$ the second Pauli matrix (for $\theta=\pi/4$
the usual Hadamard coin is recovered). As for $\hat{S}$, it just
modifies the position register one step up (down) for $s=+1(-1)$.
The last part of $\hat{U}_{\varphi}$ represents the effect of the
electric field, which adds a position linearly-dependent phase to
the states as $e^{i\varphi\hat{x}}\left\vert x,s\right\rangle =e^{i\varphi x}\left\vert x,s\right\rangle $.

It is instructive to move from position to quasi-momentum $k$ space
where $\hat{S}$ becomes diagonal with matrix elements $\exp\left\{ ik\hat{\sigma}_{3}\right\} $
with $\hat{\sigma}_{3}$ the third Pauli matrix. Hence $\hat{S}$
performs a $k$-dependent rotation of the vector state around the
$z$-axis of the Bloch sphere. For a localized initial state (which
projects on all $k$ values within the first Brillouin zone -BZ-,
$k\in\left[-\pi,\pi\right]$) this means that each of its frequency
components is rotated by a different angle. When the electric field
is added, its effect consists in introducing a $k$-dependent shift
of the quasi-momentum as \cite{Genske13} 
\begin{equation}
\exp\left(i\varphi\hat{x}\right)\left\vert \psi(k)\right\rangle =\exp\left(\varphi\partial_{k}\right)\left\vert \psi(k)\right\rangle =\left\vert \psi\left(k+\varphi\right)\right\rangle .\label{aus0}
\end{equation}

When no electric field is applied ($\varphi=0$) the problem is translational
invariant, which allows for the derivation of dispersion relations
$\omega_{\pm}(k)$, with $\omega_{+}=\omega=-\omega_{-}$ being $\omega\in\left[0,\pi\right[$
given by \cite{deValcarcel10} 
\begin{equation}
\omega=\arccos\left(\cos\theta\cos k\right),\label{reldip0}
\end{equation}
which govern the propagation of plane waves, Fig. 1(a).

Contrarily, when $\varphi\neq0$ the problem loses the translational
invariance as the unitary becomes position dependent; however, the
fact that the action of $\exp\left(i\varphi\hat{x}\right)$ is just
shifting the quasi-momentum by $\varphi$ allows one to visualize
the effect of the electric field as a shift of size $\varphi$ along
the horizontal axis, at each walk step, of the original dispersion
relations. If after $p$ steps the total accumulated phase is $q2\pi$
(with $q$ and $p$ integers) then the dispersion relation returns
to be the same as $p$ steps before, which, in \textit{continuous
}terms, suggests that the behaviour could be periodic, specially for
small $\varphi$ as the adiabaticity of the change would avoid Landau--Zener
tunneling from one to the other of the two dispersion relation curves
\cite{Regensburger11}. Indeed the presence of gaps, or pseudo-gaps,
serves as a diagnostic for localization in photonic crystals \cite{PCs}.

However, the problem is discrete, which makes of the rationality of
$\varphi/2\pi$ a crucial point, as the dispersion relation only returns
to its initial position when it is a rational $\varphi/2\pi=q/p$.

\begin{figure}

\includegraphics[scale=0.4]{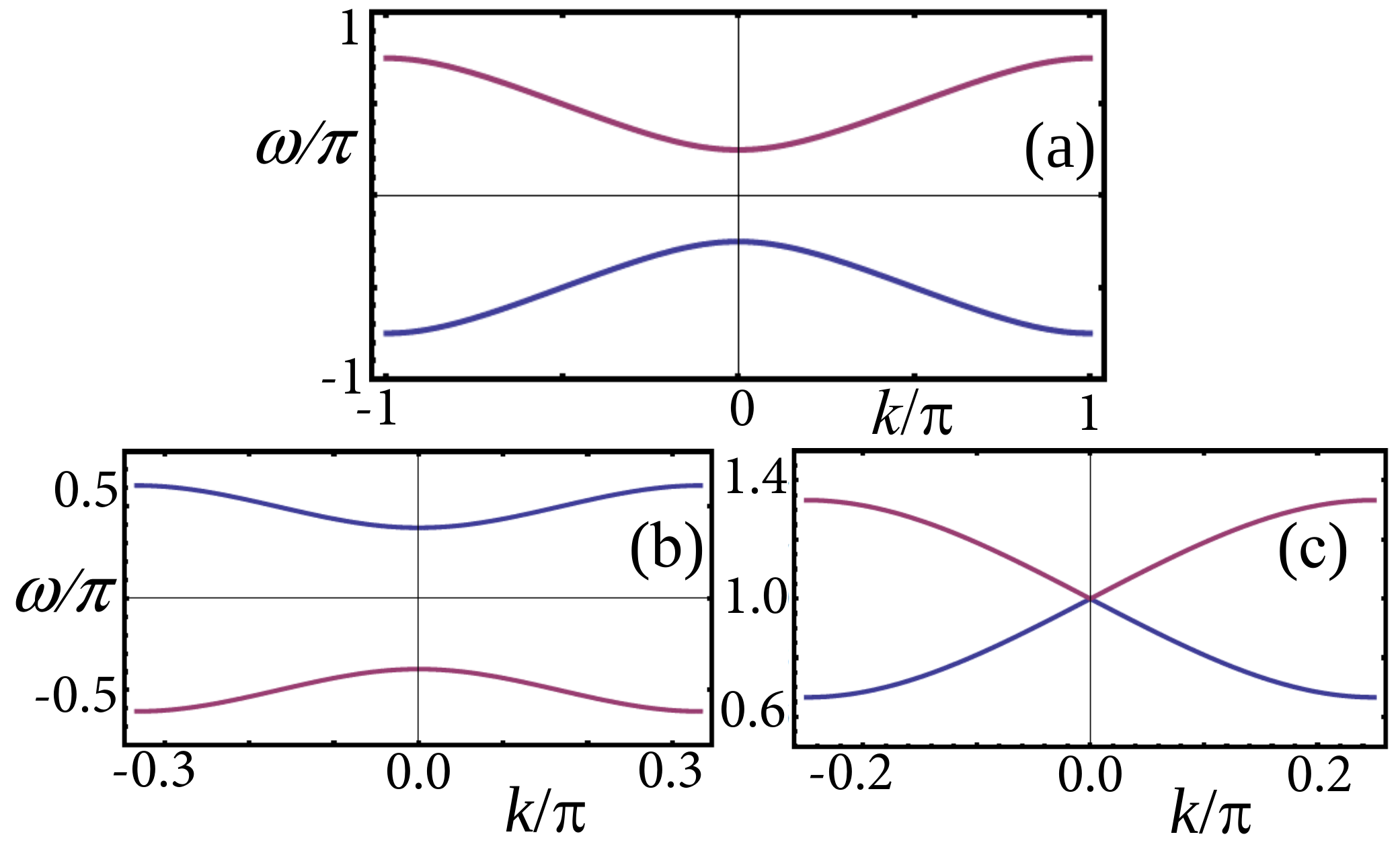}\caption{Dispersion relation (DR) of the 1D-QW for $\theta=\pi/4$ (a); effective
DRs of the 1D-EQW for $\varphi=2\pi/p$ with $p=3$ (b) and $p=4$
(c).}

\end{figure}

When $\varphi/2\pi=q/p$ there must exist a stroboscopic QW within
the EQW, which has been cleverly exploited by Cedzich \textit{et al.}
\cite{Cedzich13} in order to derive an effective dispersion relation
for the EQW. Their result reads \cite{Cedzich13} 
\begin{eqnarray}
\left.\cos\omega_{\pm}\left(k\right)\right\vert _{p}^{\left(odd\right)} & = & c_{\theta}^{p}\cos pk,\\
\left.\cos\omega_{\pm}\left(k\right)\right\vert _{p}^{\left(even\right)} & = & \left(1-c_{\theta}^{p}\right)\left(-1\right)^{1+\frac{p}{2}}-c_{\theta}^{p}\cos pk,
\end{eqnarray}
with $c_{\theta}\equiv\cos\theta$, see Fig. 1 (see Appendix A and
the treatment below of the alternate walk for details). The effective
group velocities are computed with $v_{g}\left(k\right)=p^{-1}\partial\omega\left(k\right)/\partial k$,
and it is easy to show that the maximum velocities are given by $v_{g}^{\max}=\pm c_{\theta}^{p}$
for odd $p$ (ocurring at $k=\pi/2p$), and $v_{g}^{\max}=\mp\sqrt{c_{\theta}^{p}}$
for even $p$ (ocurring at $k=0$ for $p=4n$ and $k=\pi/p$ for $p=4n+2$
with $n$ an integer). Notice that now the pseudo-momentum $k$ runs
from $-\pi/p$ to $+\pi/p$. It is clear that $v_{g}^{\max}$ rapidly
decreases as $p$ increases, hence one expects to see a transient
trapping of the probability distribution, the larger $p$ the longer
its duration, the trapping being imperfect because the maximum group
velocities, even if small, are not null. This is consistent with the
view of Landau-Zener transitions becoming less important when the
phase jumps are smaller \cite{Regensburger11} but notice that given
the discrete nature of the walk, the (transient) localization can
also occur for sizeable values of $\varphi$.

Consider now the irrational case. Any irrational number can be approximated,
with arbitrary accuracy, by the rational number obtained after a suitable
truncation of the irrational's continued fraction expansion
\begin{equation}
\frac{\varphi}{2\pi}=\left\{ a_{0},a_{1},a_{2}\ldots\right\} \equiv a_{0}+\frac{1}{a_{1}+\frac{1}{a_{2}+\ldots}}.
\end{equation}
Hence the irrational is approximated to some rational when the expansion
is finite, the closer is the approximation the more terms the expansion
contains \cite{Cedzich13}. This means that we can infer the behaviour
for the EQW in the irrational case by thinking about what occurs with
its successively more precise rational approximations. However, one
must keep in mind that in real situations (measurements, computations)
rationals and irrationals cannot be distinguished, as the number of
available digits is always finite.

Consider first the rational case $\varphi/2\pi=\left\{ 0,a_{1}\right\} =1/m_{1}$.
From the arguments above, one expects the existence of some periodicity
in the evolution of the probability distribution, and this is indeed
the case: it oscillates with a period $m_{1}$ when $m_{1}$ is odd,
and with a period $2m_{1}$ when $m_{1}$ is even (the difference
between even and odd cases lies in that the walker needs two steps
for repeating any position).

Consider now the rational $\left\{ 0,a_{1},a_{2}\right\} =n_{2}/m_{2}$
where $n_{2}=a_{2}$ and $m_{2}=1+a_{1}a_{2}$. Numerics reveal two
periods in the evolution of the distribution width, one period $m_{2}$
and also a period $m_{1}$ reminiscent of the case $\left\{ 0,a_{1}\right\} $.
For larger sequences $\left\{ 0,a_{1},a_{2},a_{3},\ldots\right\} $
the tendency is the same, with more periods that can be numerically
observed and that correspond to the periods of the subsequent possible
truncations of the sequence. Of course the weights of the different
terms $a_{n}$ determine the details of the dynamics, as periods with
very different orders of magnitude can be simultaneously involved
and, depending on the particular case, the periods are or not easily
seen.

As stated, the continued fraction expansion of an irrational number
must contain an infinite number of terms. Hence, in very general terms,
one would expect to see manifestations of all the periods (there are
an infinite number of them) contained in the successive possible truncations
of the irrational expansion, leading to a quite complicated dynamics
in general that, in any case, manifests as a permanent trapping. The
numerics confirm this and support the conclusion that the probability
distribution is really trapped for irrational values of $\varphi/2\pi$.
This is the fundamental difference between the rational and the irrational
cases: for rationals trapping is transient and lasts more the larger
$p$ is, while for irrationals the trapping is permanent. We direct
the reader to the thorough study of Cedzich et al. for full details
\cite{Cedzich13} .

\section{Electric QWs in 2D}

Unlike the 1D case, completely characterized by only four parameters
--of which only two $\left(\theta,\varphi\right)$ are dynamically
relevant--, in the 2D case the number of possible independent parameters
is actually too large to try any general treatment. Multidimensional
QWs were first introduced by Mackay \textit{et al.} \cite{Mackay02}
who generalized the 1D definition by endorsing the walker with a $2N$-dimensional
qudit, and different walks are defined depending on the $2N\times2N$
coin operator. The most widely used and studied 2D-QW is the so-called
Grover walk, whose electric version we will consider first.

In 2D-QWs the walker moves on a 2D lattice and is endorsed with a
four-dimensional coin--qudit, the evolution being governed by $\left\vert \psi\left(t+1\right)\right\rangle =\hat{S}_{2}\hat{C}_{2}\left\vert \psi\left(t\right)\right\rangle $
with $\hat{S}_{2}$ the conditional 2D-shift and $\hat{C}_{2}$ a
coin operator acting on the four--dimensional qudit. The state of
the system at (discrete) time $t$ can be written as
\begin{equation}
\left\vert \psi\left(t\right)\right\rangle =\sum_{x,y}\sum_{s}a_{x,y,s}\left(t\right)\left\vert x,y;s\right\rangle .
\end{equation}
Here $\left\vert x,y;s\right\rangle =\left\vert x,y\right\rangle \otimes\left\vert s\right\rangle $
with $\left\vert x,y\right\rangle $ the state of the walker on the
2D lattice and $\left\vert s\right\rangle $ the coin-qudit. We take
the ordering $s=X_{+},Y_{-},Y_{+},X_{-}$ for $s=1$ to $4$, respectively.
(We refer the reader to Appendix B for a brief comment about that.)
The conditional displacement in 2D reads
\begin{eqnarray}
\hat{S}_{2} & = & \sum_{x,y}\sum_{j=\pm1}[\left\vert x+j,y;X_{j}\right\rangle \left\langle x,y;X_{j}\right\vert \nonumber \\
 &  & +\left\vert x,y+j;Y_{i}\right\rangle \left\langle x,y;Y_{j}\right\vert ],
\end{eqnarray}
and the coin operator can be any unitary on the 4D coin Hilbert space.

\subsection{The Grover 2D-EQW}

For the Grover walk the coin operator matrix elements are
\begin{equation}
\left\langle j\right\vert \hat{C}_{G}\left\vert k\right\rangle =\frac{1}{2}-\delta_{jk},
\end{equation}
with $j,k=X_{\pm},Y_{\pm}$. Grover's walk dispersion relations consist
of four sheets, two of which are plane (i.e. have a null group velocity
for all $k$-values within the BZ), while the other two, non--flat,
curves cross at five conical intersections located at the center and
corners of the first BZ (see below the comments to the 2D alternate
walk). The Grover walk is well known and we refer the reader to \cite{Hinarejos13}
and references therein for further details.

We straightforwardly add the electric field to the Grover walk by
using the unitary
\begin{equation}
\hat{U}_{\varphi_{x},\varphi_{y}}=e^{i\varphi_{x}\hat{x}}e^{i\varphi_{y}\hat{y}}\hat{S}_{2}\hat{C}_{G}.
\end{equation}
Note that $e^{i\varphi_{x}\hat{x}}$ and $e^{i\varphi_{y}\hat{y}}$
commute with both $\hat{S}_{2}$ and $\hat{C}_{G}$.

While we have not been able to derive analytical results for the electric
Grover walk, we have arrived to a series of conclusions from an extensive
numerical study. First we choose the symmetric initial condition for
which the Grover and Alternate walks are equivalent \cite{DiFranco11,Hinarejos13}
(see next subsection), i.e. our initial condition does not project
onto the non-propagating sheets of the dispersion relation. This initial
state is $\left\vert c_{0}\right\rangle =2^{-1}\left(1,-1,-1,1\right)$
and in Fig. 2(a) a snapshot of the probability distribution at $t=600$
is shown. (Later on we comment on the effect of changing the initial
coin state.)
\begin{figure}
\includegraphics[scale=0.6]{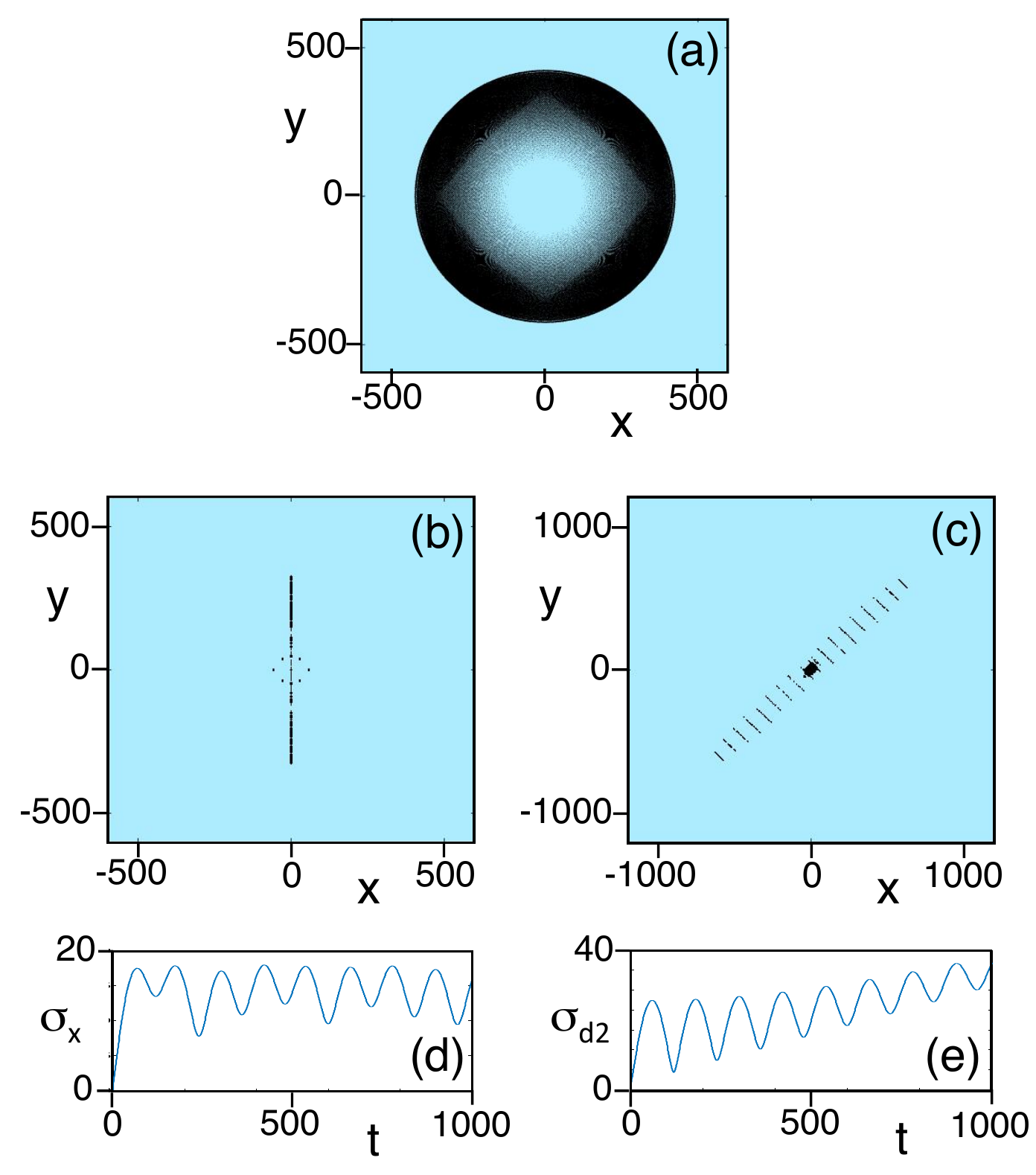}

\caption{Grover-EQW. Top view of the probability distribution in the absence
of electric field (a), and for electric fields $\varphi_{y}=0$ and
$\varphi_{x}=2\pi/120$ (b) and $\varphi_{x}=\varphi_{y}=2\pi/120$
(c); where the probabitlity is higher in the darker regions. Figs.
(d) and (e) show the evolution of the probability distribution standard
deviation along the direction in which the trapping is larger for
the cases (b) and (c), respectively. In all cases the initial coin
state is $\left\vert c\right\rangle =\left(\left\vert X_{+}\right\rangle -\left\vert Y_{-}\right\rangle -\left\vert Y_{+}\right\rangle +\left\vert X_{-}\right\rangle \right)/2=1/2\left(1,-1,-1,1\right)$
and the walker is initially located at the origin. This initial state
does not project onto the non-propagating branches of the dispersion
relation, notice the absence of localization in (a), hence the localization
effect is entirely due to the electric field. In (a) and (b) $t=600$
while in (c) $t=1000$.}

\end{figure}

From that learned in the 1D-EQW one could expect that for $\varphi_{x}=0$
and $\varphi_{y}\neq0$ (an electric field along the $y$-axis) there
could be a trapping effect, even if transient, along the $y$-axis,
but not necessarily so along the $x$-axis. The same reasoning holds
for the $\varphi_{y}=0$ case \textit{mutatis mutandi}. This is so
because the walker moves along the principal axes (the walker moves,
after one step, from (0,0) to sites $(\pm1,0)$ and $(0,\pm1)$, this
is not necessarily so for other walks), hence the directions of displacement
of the walker coincide with those of the applied electric field and
its orthogonal. This is indeed what the numerical solutions show,
see Fig. 2(b), where a snapshot of the probability distribution is
shown for $\varphi_{x}=2\pi/120$ (see caption for details), and Fig.
2(d) that shows the evolution of the distribution width (s.d.) along
the $x$-axis, $\sigma_{x}$. The figure reveals a clear localization
that can be intuitively understood as commented.

When the electric field is oriented along a line different from the
principal axes things become different. For $\varphi_{x}=\varphi_{y}=\varphi$
(i.e., an electric field directed along the principal diagonal) one
could expect two different scenarios: (i) a localization along the
principal diagonal (the direction of the field) and a spreading of
the distribution along the anti-diagonal (similarly to the axial field
case discussed above); but one could also expect (ii) a localization
effect along both the $x$ and $y$ axes (hence an effective 2D localization)
if one thinks that the projection of the field along $x$ ($y$) will
provoke a localization along $x$ ($y$). However, the fact that the
walker moves, after each step, along the principal axes excludes the
first possibility because the field direction does not coincide with
the natural directions of displacement of the walker, hence the second
option should be closer to what actually occurs.

Even if this is not entirely the case, it is quite close to it as
can be seen in Figs. 2(c) and 2(e) where we show the probability distribution
and its width for $\varphi_{x}=\varphi_{y}=2\pi/120$. It can be appreciated
how most of the probability distribution remains localized, but for
some bursts that, so to say, periodically escape from the trapping
region, the bursts being apparent in Fig. 2(c). The origin of these
imperfections in the 2D-localization can be traced back to the existence
of critical points in the Grover walk, as around these points interband
transitions cannot be avoided. We delay the discussion about that
to our study of the alternate walk below.

For other inclinations of the field we have observed complicated dynamics,
with the probability distribution usually alligning with some non-trivial
direction with respect to the electric field. See an example below,
Fig. 5, when we consider the electric alternate walk.

We must mention that for initial states other from the one we have
chosen (the one that makes the Grover walk equivalent to the alternate
walk in 2D \cite{DiFranco11}), the behaviour is qualitatively the
same, and the main difference is the existence of a more or less large
amount of trapping originated not by the electric field, but by the
projections of the initial condition onto the non-propagating branches
of the dispersion relation. Notice that these branches are not affected
by the electric field, so that the population that is trapped in those
branches would remain trapped.

As for the influence of the value of $\varphi$, we have observed
that the smaller it is the longer time the localization remains, as
in the 1D case, but we have not seen any clear difference between
rationals and ``irrationals''. This could have also been expected,
however, because in a 2D walk the trajectories that the walker can
follow are not forced to run parallel to the direction of the electric
field, as is obviously the case in the 1D walk, hence the dependence
on the size of the phase cannot be that critical as in the 1D case.

Before moving to other 2D walks, let us recall here that the Grover
walk can also be understood as the 1D walk of two independent and
non-interacting bosons. In this case, the localization along an axis
of the walk when the field is applied along that axis would mean that
the electric field acts directly only on one particle, with the result
that this particle remains at rest (during the transient localization,
of course) while the other particle on which no electric field acts,
follows a quantum walk. On the other hand, when the field acts equally
on the two particles (case $\varphi_{x}=\varphi_{y}$) we see a (transient)
co-walking of the particles as if they formed a molecule \cite{Albrecht12}.

\subsection{Electric Alternate QWs in 2D}

The Alternate Quantum Walk (AQW) was introduced as a simpler way to
build a QW on a 2D lattice using lower dimensional qudit resources
than those needed in the standard 2D walk \cite{DiFranco11,DiFranco11b,Roldan13}.

Consider a quantum walker on a 2D--lattice with a 1D internal qubit,
the basis states being $\left\vert x,y;s\right\rangle $ with $s=\pm1$.
In the electric alternate QW in 2D (EAQW) the state evolves as $\left\vert \psi\left(t+1\right)\right\rangle =\hat{U}_{\varphi_{1},\varphi_{2}}\left\vert \psi\left(t\right)\right\rangle $
where
\begin{equation}
\hat{U}_{\varphi_{1},\varphi_{2}}\equiv e^{i\left(\varphi_{x}\hat{x}+\varphi_{y}\hat{y}\right)}\hat{U}_{0,0}=e^{i\left(\varphi_{x}\hat{x}+\varphi_{y}\hat{y}\right)}\hat{S}_{y}\hat{C}_{y}\hat{S}_{x}\hat{C}_{x}.\label{eq:AEQWop}
\end{equation}
being $\hat{S}_{j}$ the conditional displacement along axis $j=x,y$,
and $\hat{C}_{j}$ given by (\ref{eq:C}) with $\alpha=\beta$, and
with, in general, different angles $\theta_{x}$ and $\theta_{y}$.
As in the electric Grover walk, we can set the two phases $\varphi_{x}$
and $\varphi_{y}$ together at the end of the operator because of
compatibility.

For $\varphi_{x}=\varphi_{y}=0$, the usual alternate QW \cite{DiFranco11,DiFranco11b}
is recovered, whose dispersion relations \cite{Roldan13} are
\begin{eqnarray}
\cos\omega_{\pm} & = & \cos\left(k_{x}+k_{y}\right)\cos\theta_{x}\cos\theta_{y}\nonumber \\
 &  & -\cos\left(k_{x}-k_{y}\right)\sin\theta_{x}\sin\theta_{y}.
\end{eqnarray}
In the usual case $\hat{C}_{x}=\hat{C}_{y}=\hat{H}$ (i.e., when the
two coins are Hadamard transformations) the dispersion relations show
a gap-less band structure with conical intersections (diabolical points),
Fig. 3(a). (The implications in the evolution of the system of these
conical intersections has been analyzed in detail in \cite{Roldan13,Hinarejos13}.)
These dispersion relations, if rotated by $\pi/4$, are the same as
the two non--plane sheets of the Grover's 2D walk dispersion relations;
hence, if an initial condition is chosen (in the Grover's walk) such
that it does not project on the plane sheets, the Grover walk and
the alternate walks become equivalent in 2D except for the $\pi/4$
rotation of the walk axes \cite{Roldan13}. This is what we did in
the previous section.

\begin{figure}
\includegraphics[scale=0.55]{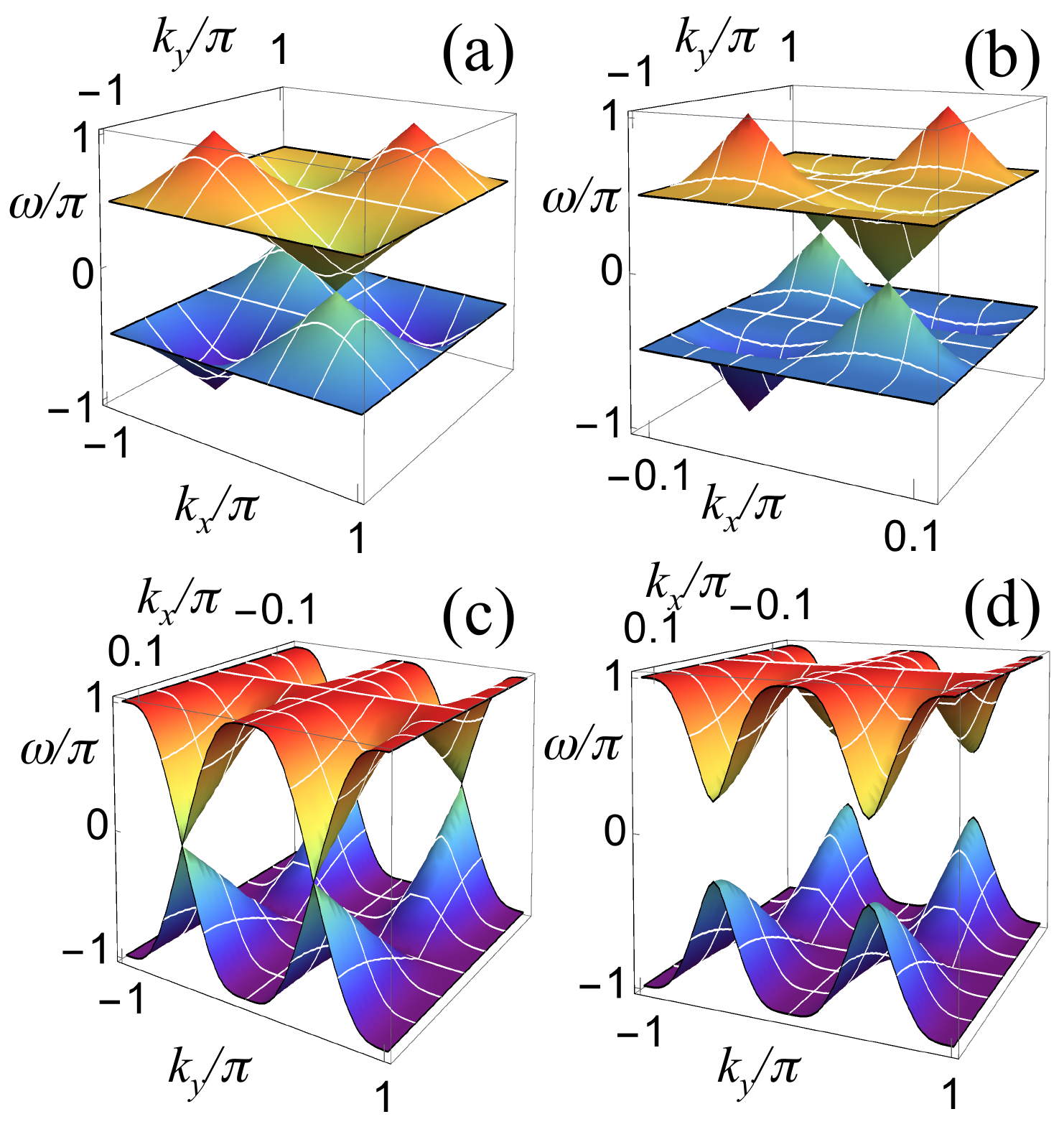}

\caption{Dispersion relations of the AQW in 2D (a), and effective dispersion
relations of the AEQW with $\varphi_{x}=2\pi/p$ and $\varphi_{y}=0$
for $p=9$ (b) and $p=8$ (c) and (d). In (b) and (c) we have taken
$\theta_{1,2}=\pi/4\pm\delta\theta$ with $\delta\theta=0$, while
in (d) $\delta\theta=0.05$.}

\end{figure}

Consider now the simplest case of an electric field directed along
one of the principal axes, say $\varphi_{y}=0$. We reason along the
same lines as in the 1D case \cite{Cedzich13} and first show that
$\hat{U}_{\varphi_{x},0}^{p}$ is translational invariant when $\varphi_{x}=2\pi q/p$:
It is easy to see that
\begin{gather}
\hat{U}_{\varphi_{x},0}\left(k_{x},k_{y}\right)\left\vert \psi\left(k_{x},k_{y}\right)\right\rangle =\nonumber \\
\hat{U}_{0,0}\left(k_{x}+\varphi_{x},k_{y}\right)\left\vert \psi\left(k_{x}+\varphi_{x},k_{y}\right)\right\rangle ,
\end{gather}
from which a similar reasoning to that in the 1D case \cite{Cedzich13}
leads to
\begin{eqnarray}
\hat{U}_{\varphi_{x},0}^{p}\left(k_{x},k_{x}\right) & = & \hat{U}_{0,0}\left(k_{x}+\varphi_{x},k_{y}\right)\nonumber \\
 &  & \hat{U}_{0,0}\left(k_{x}+2\varphi_{x},k_{y}\right)\ldots U_{0,0}\left(k_{x}+p\varphi_{x},k_{y}\right)\nonumber \\
 & = & \hat{S}_{x}\hat{U}_{0,0}\left(k_{x},k_{y}\right)\hat{S}_{x}^{2}\hat{U}_{0,0}\left(k_{x},k_{y}\right)\nonumber \\
 &  & \ldots\hat{S}_{x}^{p}\hat{U}_{0,0}\left(k_{x},k_{y}\right),
\end{eqnarray}
which has been written in a form suitable to apply the Lemma used
by Cedzich \textit{et al.} \cite{Cedzich13} (see Appendix A for details).
Finally one gets the stroboscopic dispersion relations 
\begin{eqnarray}
\left.\cos\omega_{\pm}\left(k\right)\right\vert _{p}^{\left(odd\right)} & = & \left\vert r\right\vert ^{p}\cos\left[p\arccos\left(\frac{\mathrm{Re}\left(r\right)}{\left\vert r\right\vert }\right)\right],\\
\left.\cos\omega_{\pm}\left(k\right)\right\vert _{p}^{\left(even\right)} & = & -\left\vert r\right\vert ^{p}\cos\left[p\arccos\left(\frac{\mathrm{Re}\left(r\right)}{\left\vert r\right\vert }\right)\right]+\nonumber \\
 &  & \left(-1\right)^{\frac{p}{2}+1}\left(1-\left\vert r\right\vert ^{p}\right),
\end{eqnarray}
with 
\begin{equation}
r=e^{ik_{x}}\left(c_{\theta_{x}}c_{\theta_{y}}e^{ik_{y}}-s_{\theta_{x}}s_{\theta_{y}}e^{-ik_{y}}\right),
\end{equation}
and $s_{\theta_{x}}=\sin\theta_{x}$, etc. From $\omega_{\pm}\left(k\right)$
the group velocities are easily obtained as $v_{g}=(1/p)\partial\omega/\partial k$,
we do not give them because of their length.

In Figs. 3 we represent these effective dispersion relations for $p=9$,
Fig. 3(b), and $p=8$, Figs. 3(c) and (d); where we took $\theta_{x,y}=\pi/4\pm\delta\theta$
with $\delta\theta=0$ in (b) and (c), and $\delta\theta=0.05$ in
(d). We have chosen small values for $p$ in order to easily visualize
the curves, as the slopes quickly increase as $p$ does, and for larger
$p$ it is difficult to see the details. We see that the conical intersections
of the original dispersion relation survive with the electric field,
and we see that a non--null $\delta\theta$ breaks these degeneracies
\cite{Roldan13}. Hence one must expect large changes in the dynamics
depending on whether $\delta\theta=(\theta_{x}-\theta_{y})/2$ is
null or not. In any case, notice that for most $k$ values there are
wide plateaus in the effective dispersion relations in which the group
velocities are relatively very small. Moreover, the plateaus become
wider as $p$ is increased, so that the smallness of the group velocity
across most of the $k$-space is a most remarkable feature. Notice
finally that $k_{y}$ now runs from $-\pi/2p$ to $\pi/2p$. Apart
from these very general considerations, not much information can be
extracted from the dispersion relations when the initial state is
localized in space, as it is our present case, as the effect of interferences
on the propagation of each frequency component cannot be easily devised.

In order to study numerically the behaviour of the EAQW it is convenient
to make use of the partial equivalence with the Grover walk, keeping
in mind the $\pi/4$ rotation between the two walks. This rotation
reflects the fact that the directions of displacement of the walker
do not coincide with the principal axes, but with the diagonals, as
at each step the walker is displaced along both $x$ and $y$. This
implies that when a field is added along, say, $y$, this actually
means that the walker acquires a phase $\varphi$ at both $x$ and
$y$ displacements, which is equivalent to the action of a field of
amplitude $2\varphi$ directed along the main diagonal in the electric
Grover walk.

The $\pi/4$ rotation manifests in the numerical results: The plots
shown in Fig. 2 above for the Grover walk with fields of amplitude
$\varphi$ along $x$, in Fig. 2(b), and along the main diagonal,
in Fig. 2(c), apply to the EAQW with $2\varphi$ fields along the
diagonal and along the $x$-axis, respectively, if rotated 45 degrees.

However the alternate walk allows one extra degree of freedom, because
when $\theta_{x}$ is taken different from $\theta_{y}$, the conical
degeneracies are removed, which allows to see their influence

\begin{figure}
\includegraphics[scale=0.45]{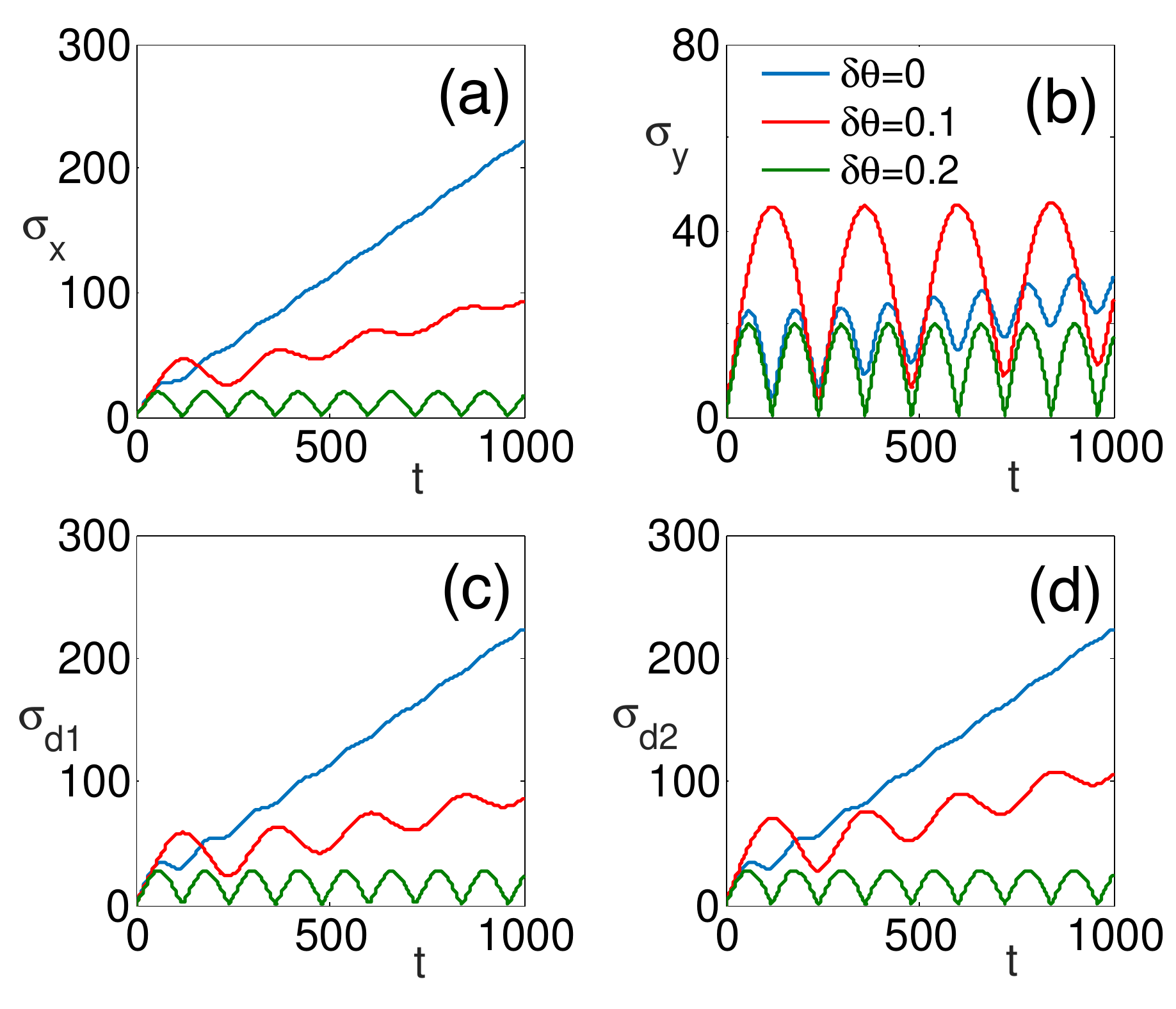}

\caption{AEQW. Evolution of the probability distribution width along the principal
axes, (a) and (b), and the diagonals, (c) and (d), for $\varphi_{x}=2\pi/120$
and $\varphi_{y}=0$. In all cases three different values of $\delta\theta$
($=0$, $0.1$, and $0.2$; curves with higher to lower values at
$t=1000$) have been considered, where $\theta_{x,y}=\pi/4\pm\delta\theta$.}

\end{figure}

In Figs. 4 we represent the evolution of the distribution width, along
both the principal axes and the diagonals, for three values of $\delta\theta$
and for a field $\varphi_{x}=2\pi/120$ and $\varphi_{y}=0$. The
case $\delta\theta=0$ reveals that there is a transient trapping
along the $y$-axis. When the conical intersections are lifted for
$\delta\theta\neq0$ the progressive appearance of a perfect trapping
along all four axes is clear. Again, this is consistent with the view
that Landau-Zener transitions become less likely the larger is the
energy gap.

We have also investigated the effect of tilted fields, and in Fig.
5 we show results for the case $\varphi_{x}=2\varphi_{y}=2\pi/120$.
The figure reveals that also in this case there occurs a perfect trapping
when $\delta\theta$ is large enough, which emphasizes the role played
by the conical intersections.

\begin{figure}
\includegraphics[scale=0.7]{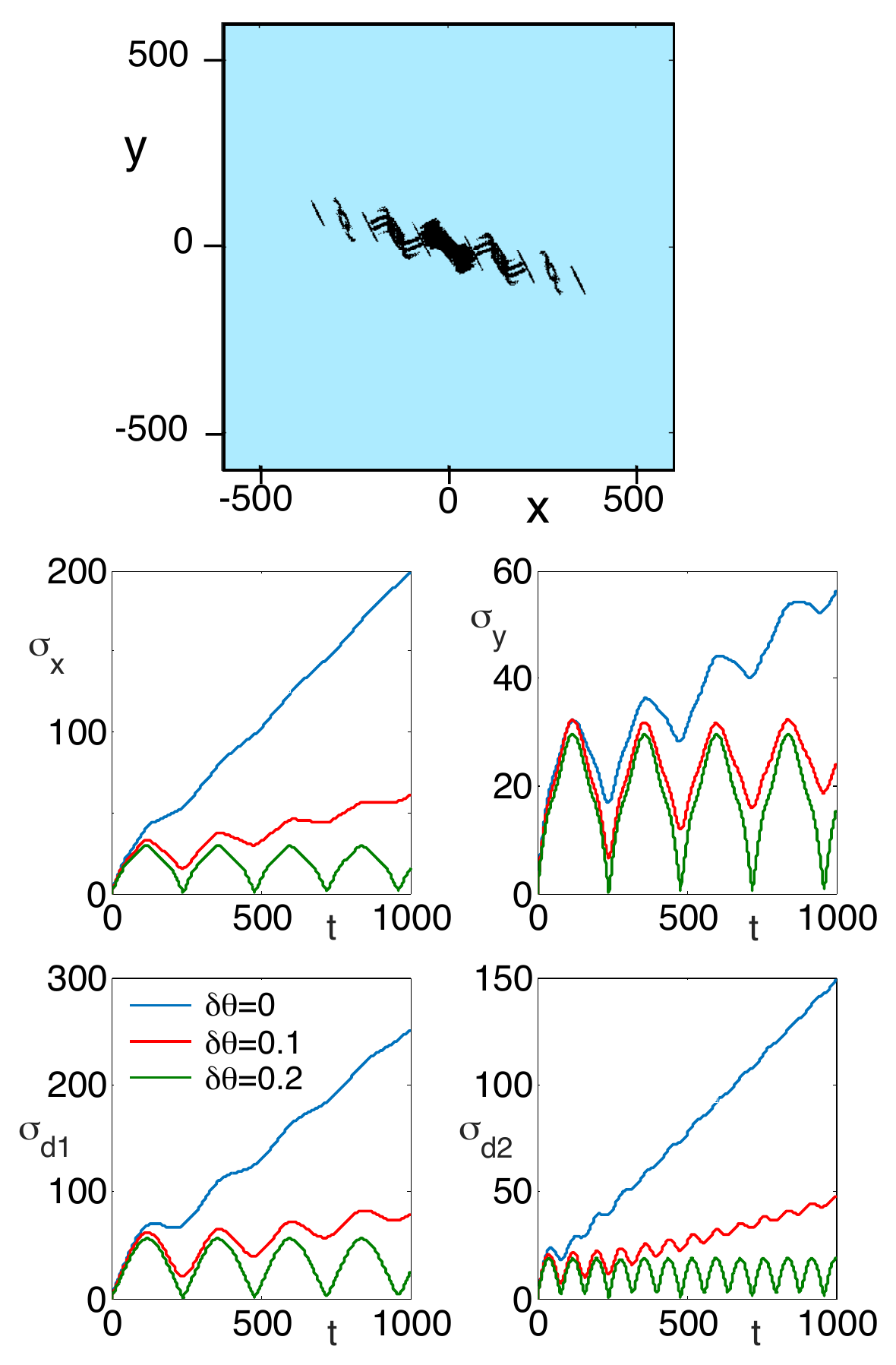}

\caption{AEQW. Top figure is a snapshot of the probability distribution for
$\varphi_{x}=2\varphi_{y}=2\pi/120$ at $t=600$. The rest of the
figures show, as in Fig. 4, the evolution of the probability distribution
width.}

\end{figure}

\subsection{Other EQWs in 2D: DFT and Hadamard coins}

As deriving general analytical results with electric fields is beyond
reach in this case, looking at the walk dispersion relation seems
the obvious simple choice, as if they contain diabolical points one
can expect a behaviour similar to that of the Grover walk, while a
perfect 2D trapping for some appropriate field direction could be
expected when there are no conical intersections.

Let us consider first the 2D-DFT walk, whose coin reads

\begin{equation}
\hat{C}_{DFT}=\frac{1}{2}\left(\begin{array}{cccc}
1 & 1 & 1 & 1\\
1 & i & -1 & -i\\
1 & -1 & 1 & -1\\
1 & -i & -1 & i
\end{array}\right).
\end{equation}
The dispersion relations are not explicit and are obtained by numercically
solving
\begin{gather}
\cos2\omega+2\sin2\omega=\cos\left(k_{x}-k_{y}\right)+\nonumber \\
2\left(\sin\omega+\sin k_{x}+\sin k_{y}+\cos k_{x}+\cos k_{y}\right)\sin\omega
\end{gather}
see Fig. 6. It consists of four sheets with several conical intersections
and a lesser degree of symmetry than the Grover walk. In Fig. 6(b)
we have represented the probability distribution at $t=600$, see
caption.

The effect of the electric field on the DFT-walk is similar to that
in the Grover walk, at least qualitatively, as it can be seen by comparing
Figs. 7 and 2. Clearly the trapping capacity when the field is along
the diagonal is much smaller in the DFT coin. Hence, we conclude that
the electric-DFT walk is very similar to the electric Grover-walk,
the differences beeing more quantitative than qualitative.

\begin{figure}
\includegraphics[scale=0.32]{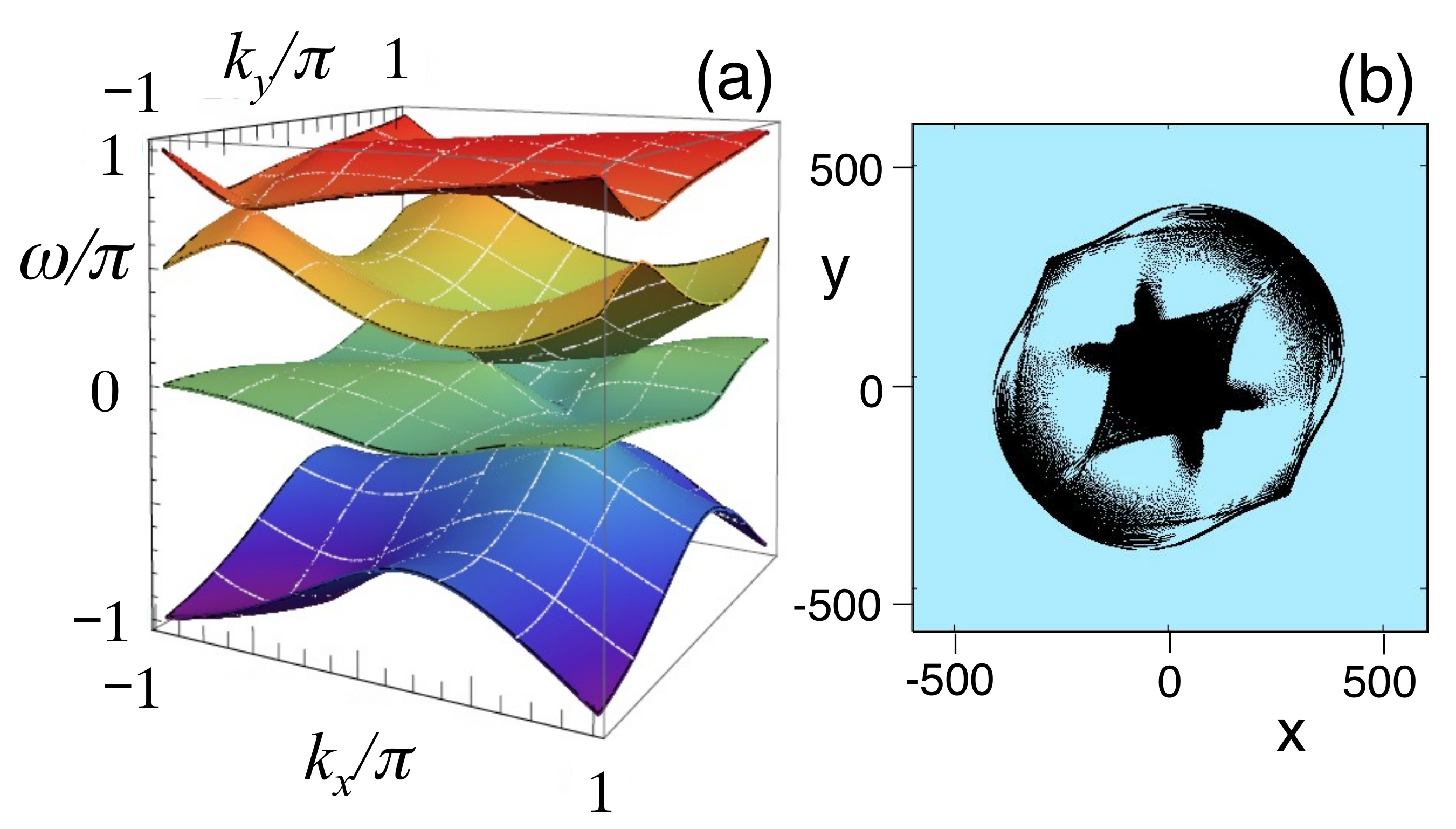}

\caption{Dispersion relation of the DFT-QW in 2D (a), and a snapshot of the
probability distribution at $t=600$ (b). The initial coin state is
$\left\vert c\right\rangle =1/2\left(1,i,i,-1\right)$.}

\end{figure}

\begin{figure}
\includegraphics[scale=0.55]{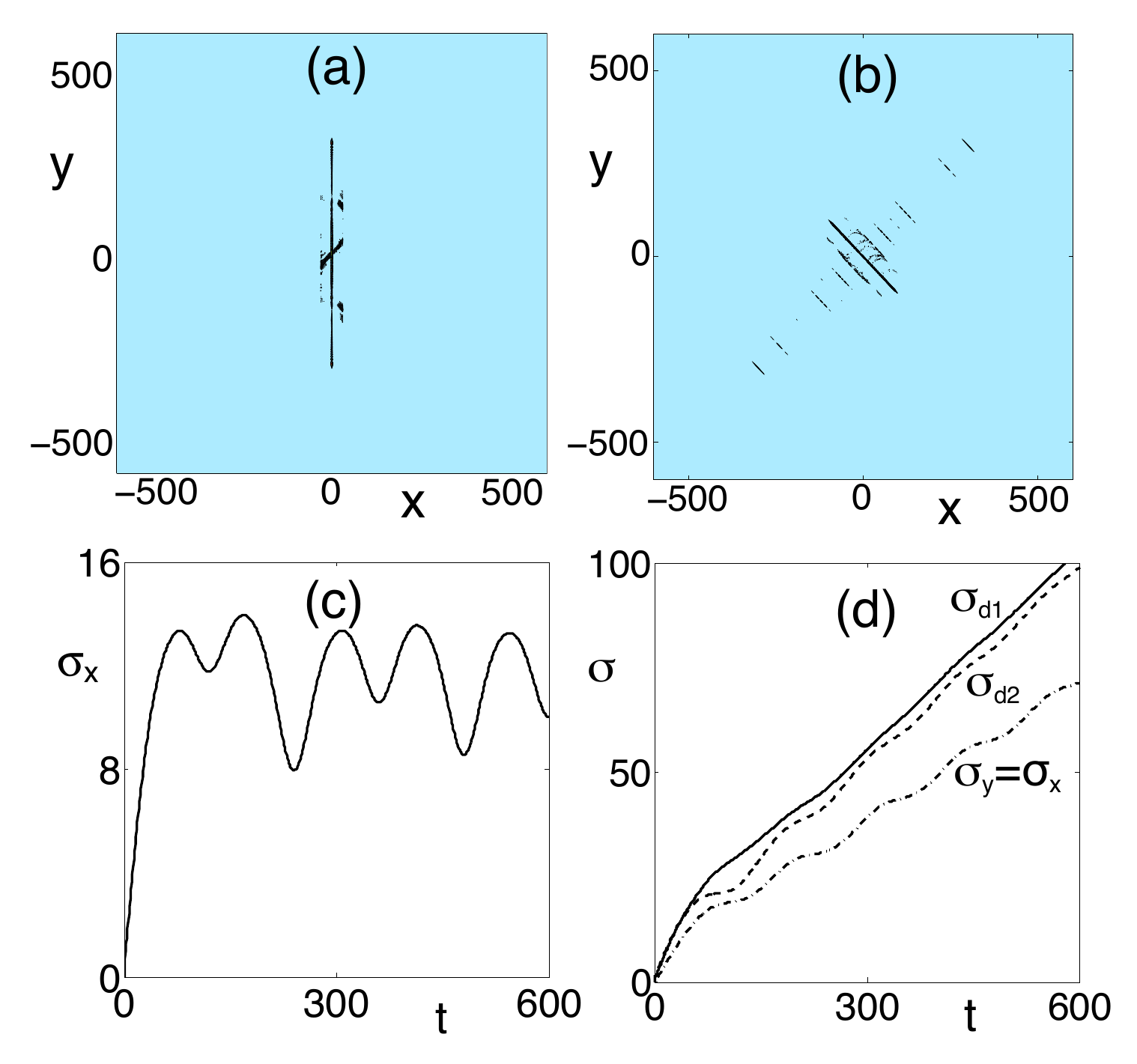}

\caption{Electric DFT walk in 2D. In Figs. (a) and (c) the electric field is
$\varphi_{x}=2\pi/120$ and $\varphi_{y}=0$, whilst in Figs. (b)
and (d) the electric field is $\varphi_{x}=\varphi_{y}=2\pi/120$.
The initial state is as in Fig. 6 and $t=600$.}

\end{figure}

Let us finally consider the Hadamard walk, whose coin is defined as
$\hat{C}_{H_{2}}=\hat{H}\otimes\hat{H}$, i.e. 
\begin{equation}
\hat{C}_{H_{2}}=\frac{1}{2}\left(\begin{array}{cccc}
1 & 1 & 1 & 1\\
1 & -1 & 1 & -1\\
1 & 1 & -1 & -1\\
1 & -1 & -1 & 1
\end{array}\right),\label{Hadamard}
\end{equation}
which is a separable coin when acting on vector $\left\vert s\right\rangle =\mathrm{col}(X_{+},Y_{-},Y_{+},X_{-})$
\cite{Mackay02} (see Appendix B). Its dispersion relation reads \cite{Annabestani10}
\begin{eqnarray}
\cos\omega_{1,2} & = & \frac{1}{4}\left[\cos k_{x}-\cos k_{y}\mp\sqrt{\mathcal{S}+8}\right],\\
\mathcal{S} & = & \cos^{2}k_{x}+\cos^{2}k_{y}+6\cos k_{x}\cos k_{y},
\end{eqnarray}
and its four sheets can be seen in Fig. 8, that appear distributed
by pairs that intersect along a line within each pair, but the two
pairs are well separated by a clear gap.

\begin{figure}
\includegraphics[scale=0.35]{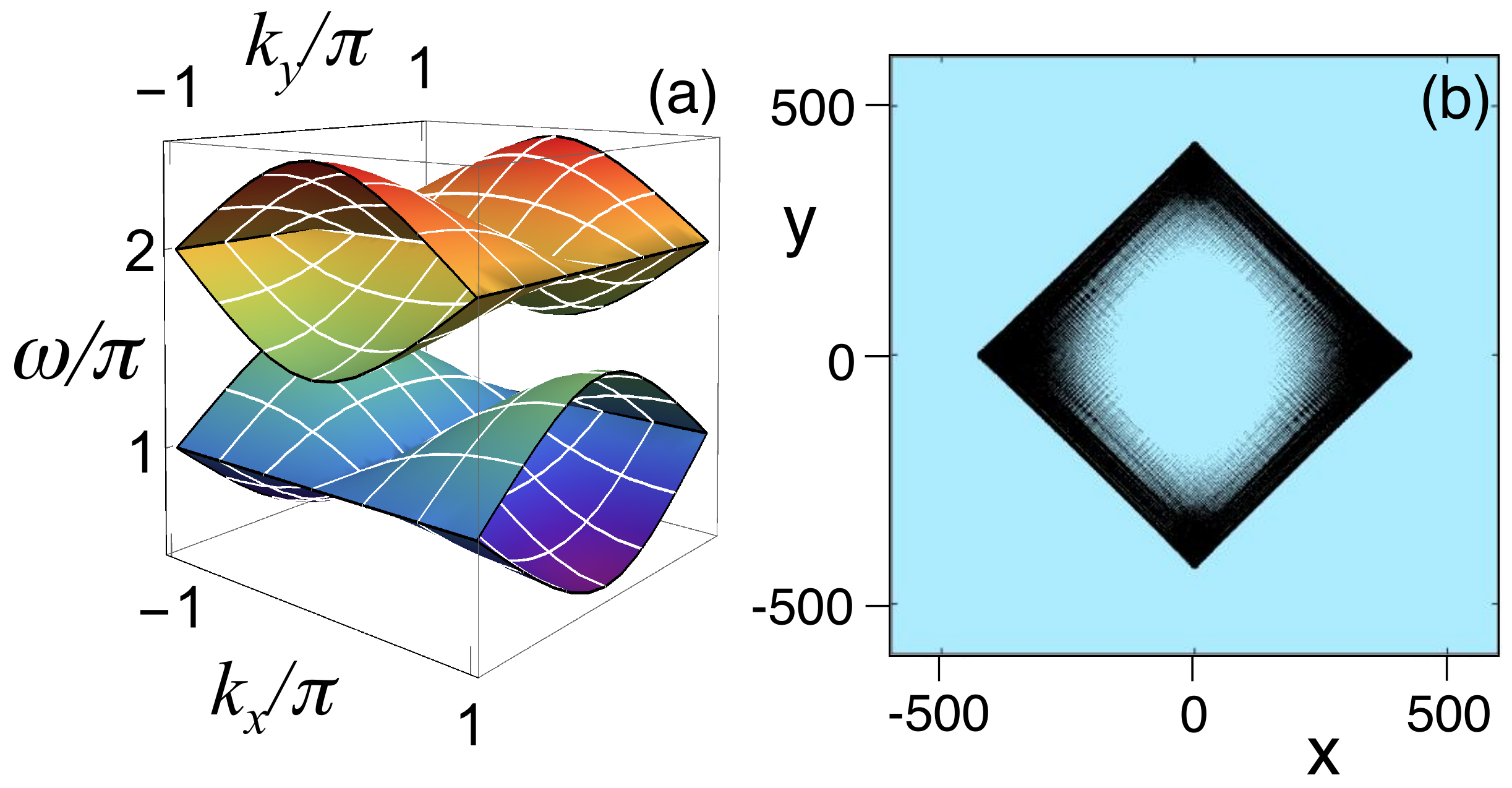}

\caption{Dispersion relation of the Hadamard-QW in 2D (a), and snapshot of
the probability distribution at $t=600$ (b). The initial coin state
is $\left\vert c\right\rangle =1/2\left(1,i,i,-1\right)$.}

\end{figure}

The effect of the electric field on this walk is sharper than in any
of the previous ones but the alternate walk with different angles
in the two coin operations. Now we see clear 1D or 2D trapping depending
on the direction of the field as Figs. 9 and 10 illustrate. Figs.
9 show the behaviour for a field directed along the main diagonal,
which produces a sharp diagonal trapping, while in Fig. 10 it is shown
how a field directed along the $x$-axis produces a sharp 2D trapping.
The behaviour reveals that in this Hadamard walk the main directions
of propagation do not coincide with those of the main axes. The 2D-Hadamard
electric walk confirms the role played by conical intersections in
the dispersion relations leading to the conclussion that perfect 2D
trapping can occur if such diabolical points are removed.

\begin{figure}
\includegraphics[scale=0.5]{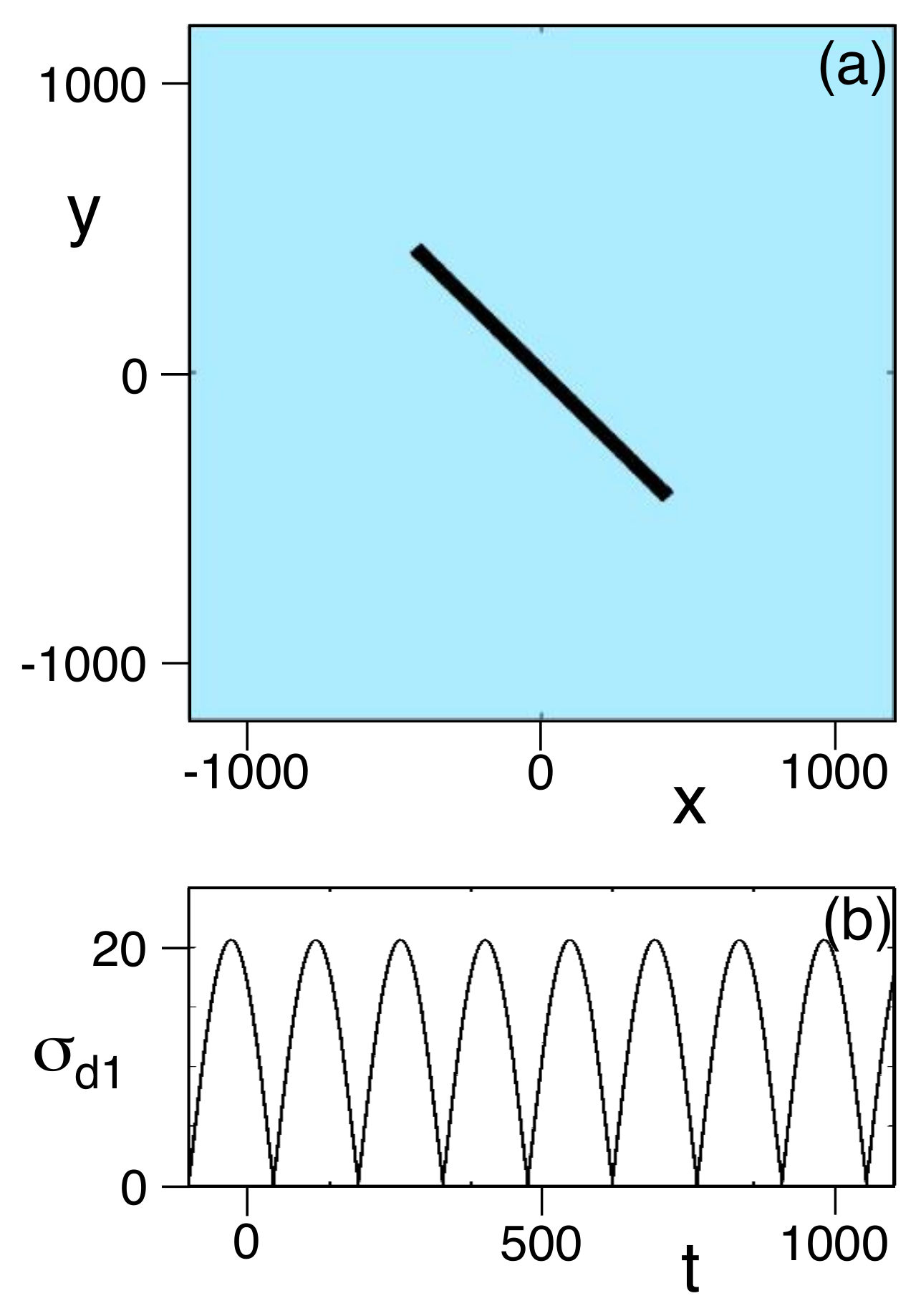}

\caption{Electric Hadamard-QW in 2D when the electric field is $\varphi_{x}=2\pi/120$
and $\varphi_{y}=0$. (a) shows a snapshot of the probability distribution
at $t=1000$, and (b) the evolution of its width. The initial coin
state is $\left\vert c\right\rangle =1/2(1,i,i,-1)$.}

\end{figure}

\begin{figure}
\includegraphics[scale=0.4]{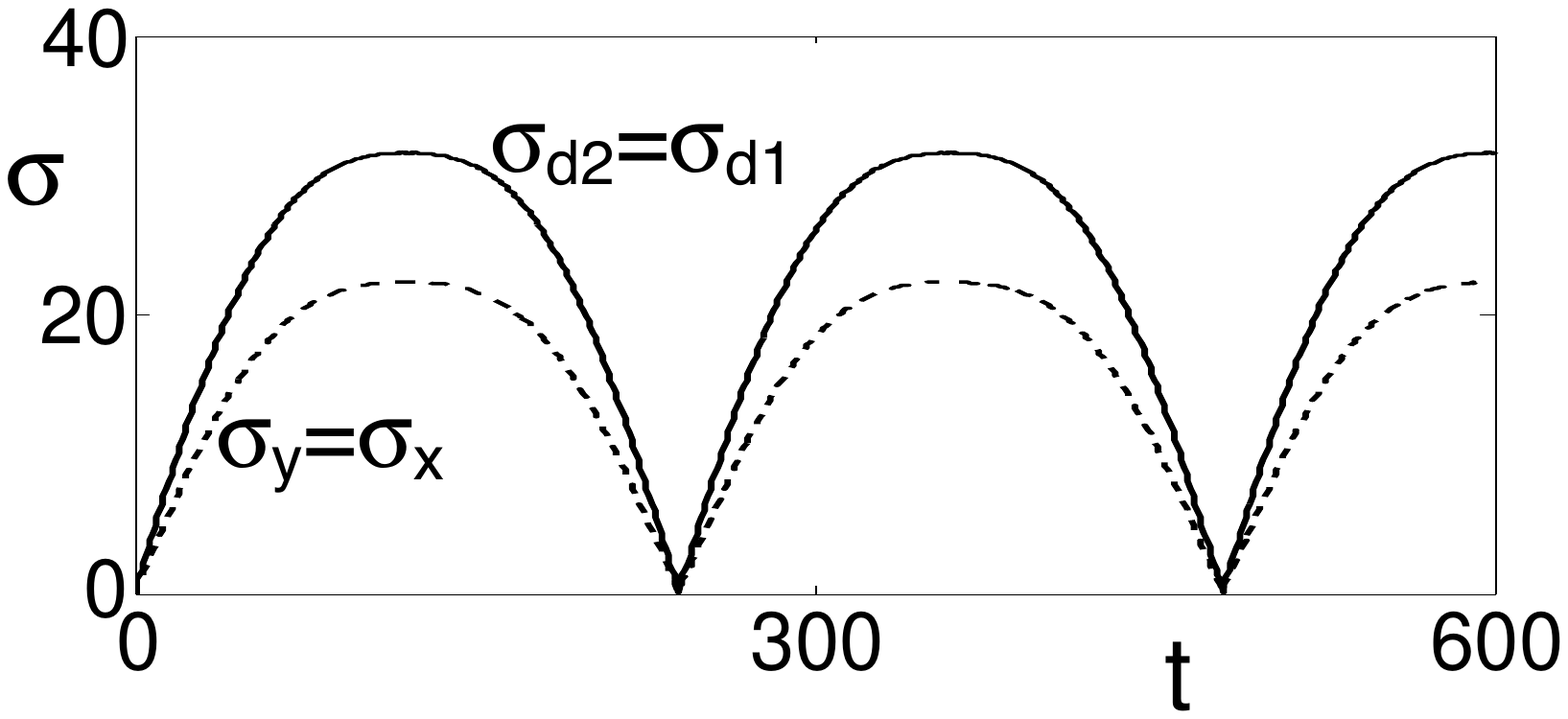}

\caption{Evolution of the probability distribution width for the electric Hadamard-QW
in 2D when the electric field is $\varphi_{x}=\varphi_{y}=2\pi/120$.
A perfect 2D trapping is apparent.}

\end{figure}

\section{Conclusions}

After a suitable review of the one dimensional case, we have addressed
the study of electric quantum walks in two dimensions. The easiest
one to understand in intuitive terms is the Grover walk: a field along
the principal axes produces trapping along the field axis, and when
the field acts along the diagonals, a frustrated 2D trapping occurs.
The reason for the frustration lies in the existence of conical intersections
in the Grover walk dispersion relations, which we have confirmed through
the analysis of the alternate electric QW. As the two walks are equivalent
after a $\pi/4$ rotation of the principal axes, the behaviour of
the alternate electric walk is somewhat counter-intuitive at first
sight as the field seems to be rotated by $\pi/4$. We have also addressed
the two other popular 2D walks, namely the Hadamard and DFT walks.

In our analysis of the alternate walk we have shown that when the
two coin operations within each step are not identical, which removes
conical intersections, the effect of the electric field can lead to
a perfect 2D localization, a result confirmed by the behaviour of
the Hadamard walk that exhibits perfect 2D trapping for appropriate
field direction.

\textbf{Acknowledgements}

We gratefully acknowledge continued financial support from Spanish
Government (Ministerio de Educación e Innovación) through projects
FIS2011-26960, FIS2014-60715-P, FPA2011-23897, FPA2014-54459-P, and
SEV-2014-0398, and support from the Generalitat Valenciana (Valencian
Goverment) through\ grant GV-PROMETEO-II-2014-087. MH acknowledges
financial support from ANII (Uruguay) grant PD\_NAC\_2014\_1\_102359.

\section{Appendix A}

Cedzich \textit{et al.} \cite{Cedzich13} introduce the following
Lemma. Let $m\in\mathbb{N}$ and let $\eta$ be a primitive $m$-th
root of unity. Now consider the matrices 
\begin{equation}
\hat{A}=\left(\begin{array}{cc}
a & b\\
c & d
\end{array}\right),\ \ \hat{R}=\left(\begin{array}{cc}
\eta & 0\\
0 & \eta^{-1}
\end{array}\right),
\end{equation}
and set
\begin{equation}
\tau_{m}=Tr\left[\hat{A}\hat{R}^{0}\hat{A}\hat{R}^{1}\ldots\hat{A}\hat{R}^{m-1}\right],
\end{equation}
then it can be demonstrated that \label{Lemma 1} 
\begin{eqnarray}
\tau_{m}^{\left(odd\right)} & = & a^{m}+d^{m},\\
\tau_{m}^{\left(even\right)} & = & -\left(a^{m}+d^{m}\right)+2\left(-1\right)^{m/2}\nonumber \\
 &  & \times\left[\left(ad\right)^{m/2}-\left(\det A\right)^{m/2}\right].
\end{eqnarray}

In order to apply the Lemma we need the matrix elements of $\hat{U}_{0,0}\left(k_{x},k_{y}\right)$,
wich can be written as 
\begin{gather}
\hat{U}_{0,0}\left(k_{x},k_{y}\right)=\left(\begin{array}{cc}
a & b\\
-b^{\ast} & a^{\ast}
\end{array}\right),\\
a=e^{i\left(k_{x}+k_{y}\right)}\cos\theta_{x}\cos\theta_{y}-e^{i\left(k_{x}-k_{y}\right)}\sin\theta_{x}\sin\theta_{y},\\
b=e^{i\left(k_{x}+k_{y}\right)}\sin\theta_{x}\cos\theta_{y}+e^{i\left(k_{x}-k_{y}\right)}\cos\theta_{x}\sin\theta_{y},
\end{gather}
and then from the trace $\tau_{p}=a^{p}+\left(a^{\ast}\right)^{p}$
one gets the dispersion relations 
\begin{eqnarray}
\left.\cos\left[\omega_{\pm}\left(k\right)\right]\right\vert _{p}^{\left(odd\right)} & = & \left\vert a\right\vert ^{p}\cos\beta,\\
\left.\cos\left[\omega_{\pm}\left(k\right)\right]\right\vert _{p}^{\left(even\right)} & = & \left(-1\right)^{\frac{p}{2}+1}\left(1-\left\vert a\right\vert ^{p}\right)-\left\vert a\right\vert ^{p}\cos\beta,\qquad\\
\beta & \equiv & p\arccos\left(\frac{\mathrm{Re}\left(a\right)}{\left\vert a\right\vert }\right).
\end{eqnarray}

\section{Appendix B}

Multidimensional QWs can be defined in different ways. In the original
definition \cite{Mackay02} the ordering of the coin elements is taken
from the tensor product of two qubits: if $\mathrm{col}\left(x_{+},x_{-}\right)$
and $\mathrm{col}\left(y_{+},y_{-}\right)$ are these qubits, then
the tensor product qudit has elements $\mathrm{col}\left(x_{+}y_{+},x_{+}y_{-},x_{-}y_{+},x_{-}y_{-}\right)$.

It is natural to interpret $x_{+}y_{+}$ ($x_{-}y_{-}$) as the positive
(negative) direction of the main diagonal and $x_{+}y_{-}$ ($x_{-}y_{+}$)
as the negative (positive) direction of the antidiagonal, so that
one can write the coin vector components as $\mathrm{col}\left(d_{+},a_{-},a_{+},d_{-}\right)$
with $d$ ($a$) standing for diagonal (antidiagonal). This is the
ordering chosen by, e.g., \cite{Mackay02}. However the above choice
seems to leave unoccupied positions on the plane, and it is also natural
to avoid this by rotating clockwise the displacement directions by
45º, so that the coin vector is written as $\mathrm{col}\left(x_{+},y_{-},y_{+},x_{-}\right)$,
which is the choice made, e.g., by \cite{Annabestani10} and that
we have also followed. Of course the two choices are equivalent, but
for a 45º rotation.

But the above orderings are not mandatory and one can order the coin
components at will. For example, in \cite{Hinarejos13} we chose the
ordering $\mathrm{col}\left(x_{+},x_{-},y_{+},y_{-}\right)$, mainly
because it admits a straightforward notation in the generalization
to higher dimensional QWs.

This can be confusing, of course, as authors do not use to specify
which ordering have they chosen, which makes sometimes difficult to
translate the results of different articles. For example, if one uses
the definitions of \cite{Mackay02,Annabestani10}, the Hadamard coin
expression is that given in the main text, Eq. (\ref{Hadamard}),
that is designed to be a separable matrix, but if the same matrix
is applied to other ordering, e.g. ours in \cite{Hinarejos13}, then
one is not implementing a separable operator, and one is indeed making
a different QW. In fact, if one is willing to implement the Hadamard
walk with the ordering of the coin elements in \cite{Hinarejos13},
the correct matrix must be written as  
\begin{equation}
\hat{C}_{H^{\prime}}=\frac{1}{2}\left(\begin{array}{cccc}
1 & 1 & 1 & 1\\
1 & 1 & -1 & -1\\
1 & -1 & -1 & 1\\
1 & -1 & 1 & -1
\end{array}\right),
\end{equation}
which is obtained by permuting the second and fourth indexes of matrix
(\ref{Hadamard}), as this is the difference between the two orderings
we have discussed. It is fortunate the fact that the Grover coin has
the same expression irrespectively of the ordering chosen for the
coin vector elements.


\begin{thebibliography}{10}
\bibitem{Kempe03} J. Kempe, Contemp. Phys. \textbf{44}, 307 (2003).

\bibitem{Kendon06} V. Kendon, Math. Struct. Comput. Sci. \textbf{17},
1169 (2006); Phil. Trans. R. Soc. A \textbf{364}, 3407 (2006).

\bibitem{Konno08} N. Konno, \textit{Quantum Walks in Quantum Potential
Theory} (Lecture Notes in Mathematics vol 1954) ed U Franz and M Schürmann
(Berlin: Springer, 2008) pp 309--452.

\bibitem{Venegas12} S.E. Venegas-Andraca, Quant. Inf. Proc. \textbf{11},
1015 (2012).

\bibitem{Strauch06} F. W. Strauch, Phys. Rev. A 73, 054302 (2006).

\bibitem{deValcarcel10} G.J. de Valcárcel, E. Roldán, and A. Romanelli,
New J. Phys \textbf{12}, 123022 (2010).

\bibitem{Hinarejos13} M. Hinarejos, A. Pérez. E. Roldán, A. Romanelli,
and G.J. de Valcárcel, New J. Phys. \textbf{15}, 073041 (2013).

\bibitem{Strauch07}F. W. Strauch, J. Math. Phys. \textbf{48}, 082102
(2007).

\bibitem{Chandrasekar13} C.M. Chandrasekar, Sci. Reports Sci. Rep.
\textbf{3}, 2829 (2013).

\bibitem{DiMolfetta14} G. Di Molfetta, M. Brachet, and F. Debbasch,
Physica A \textbf{397}, 157 (2014).

\bibitem{Aharonov93} Y. Aharonov, L. Davidovich, and N. Zagury, Phys.
Rev. A \textbf{48}, 1687 (1993).

\bibitem{Meyer96} D. Meyer, J. Stat. Phys. \textbf{85}, 551 (1996).

\bibitem{Farhi98} E. Farhi and S. Gutmann, Phys. Rev. A \textbf{58},
915 (1998).

\bibitem{Childs04} A.M. Childs and J. Goldstone, Phys. Rev. A \textbf{70},
042312 (2004).

\bibitem{Feynman65} Richard P. Feynman and A.R. Hibbs, \textit{Quantum
Mechanics and Path Integrals}, International Series in Pure and Applied
Physics (McGraw-Hill, 1965).

\bibitem{Watrous01} J. Watrous, Proceedings of the 33rd Annual ACM
Symposium on the Theory of Computing, Heraklion, Crete, Greece, July
06--08, 2001 (ACM Press, New York, 2001), p.60.

\bibitem{Childs09} A.M. Childs, Phys. Rev. Lett. \textbf{102}, 18050
(2009).

\bibitem{Lovett10} N.B. Lovett, S. Cooper, M. Everitt, M. Trevers,
and V. Kendon, Phys. Rev. A \textbf{81}, 042330 (2010).

\bibitem{review exp} K. Manouchehri and J. Wang, \textit{Physical
Implementation of Quantum Walks }(Springer, 2014).

\bibitem{Bouwmeester99} D. Bouwmeester, I. Marzoli, G. P. Karman,
W. Schleich, and J. P. Woerdman, Phys. Rev. A \textbf{61}, 013410
(1999).

\bibitem{Knight03} P. L. Knight, E. Roldán, and J. E. Sipe, Phys.
Rev. A \textbf{68}, 020301 (2003); Opt. Commun. \textbf{227}, 147
(2003).

\bibitem{Roldan05} E. Roldán and J. C. Soriano, J. Mod. Opt. \textbf{52},
2649 (2005).

\bibitem{magnetic QW} P. Arnault and F. Debbasch, Physica A \textbf{443},
179 (2016).

\bibitem{magnetic2} I. Yalç\i{}nkaya and Z. Gedik, Phys. Rev. A \textbf{92},
042324 (2015).

\bibitem{Wojcik04} A. Wojcik, T. Lukzak, P. Kurzynski, A. Grudka,
and M. Bednarska, Phys. Rev. Lett. \textbf{93}, 180601 (2004).

\bibitem{Meyer97}D. A. Meyer, Phys. Rev. E \textbf{55}, 5261 (1997).

\bibitem{Meyer98}D. A. Meyer, J. Phys. A: Math. Gen. \textbf{31},
2321 (1998).

\bibitem{Banuls06} M.C. Bañuls, C. Navarete, A. Pérez, E. Roldán,
and J.C. Soriano, Phys. Rev. A \textbf{73}, 062304 (2006).

\bibitem{Regensburger11} A. Regensburger, Ch. Bersch, B. Hinrichs,
G. Onishchukov, A. Schreiber, Ch. Silberhorn, and U. Peschel, Phys.
Rev. Lett. \textbf{107}, 233902 (2011).

\bibitem{Cedzich13} C. Cedzich, T. Rybár, A. H. Werner, A. Alberti,
M. Genske, and R. F. Werner, Phys. Rev. Lett. \textbf{111}, 160601
(2013).

\bibitem{Genske13} M. Genske, W. Alt, A. Steffen, A. H. Werner, R.
F. Werner, D. Meschede, and A. Alberti, Phys. Rev. Lett. \textbf{110},
190601 (2013).

\bibitem{Xue15} P. Xue, R. Zhang, H. Qin, X. Zhan, Z.H. Bian, J.
Li, and B.C. Sanders, Phys. Rev. Lett. \textbf{114}, 140502 (2015).

\bibitem{Linden09} N. Linden and J. Sharam, Phys. Rev. A \textbf{80},
052327 (2009).

\bibitem{Shikano10} Y. Shikano and H. Katsura, Phys. Rev. E \textbf{82},
031122 (2010).

\bibitem{Hofstadter} D. R. Hofstadter, Phys. Rev. B \textbf{14},
2239 (1976).

\bibitem{DiFranco15} C. Di Franco and M. Paternostro, Phys. Rev.
A \textbf{91}, 012328 (2015).

\bibitem{PCs} S. John , Phys. Rev. Lett. \textbf{58}, 2486 (1987);
E. Yablonovitch, Phys. Rev. Lett. \textbf{58}, 2059 (1987).

\bibitem{Mackay02} T.D.Mackay, S.D. Bartlett,L.T.Stephenson, and
B. C. Sanders, J. Phys. A: Math. Gen. \textbf{35}, 2745 (2002).

\bibitem{Albrecht12} A. Ahlbrecht, A. Alberti, D. Meschede, V.B.
Scholz, A.H. Werner, and R.F. Werner, New J. Phys. \textbf{14}, 073050
(2012).

\bibitem{DiFranco11} C. Di Franco, M. Mc Gettrick, and Th. Busch,
Phys. Rev. Lett. \textbf{106}, 080502 (2011).

\bibitem{DiFranco11b} C. Di Franco,M.Mc Gettrick, T. Machida, and
Th. Busch, Phys. Rev. A \textbf{84}, 042337 (2011).

\bibitem{Roldan13} E. Roldán, C. Di Franco, F. Silva, and G.J. de
Valcárcel, Phys. Rev. A \textbf{87}, 022336 (2013).

\bibitem{Kollar14} B. Kollár, J. Novotný, T. Kiss, and I. Jex, New
J. Phys. \textbf{16}, 023002 (2014).

\bibitem{Annabestani10} M. Annabestani, M. R. Abolhasani, and G.
Abal, J. Phys. A: Math. Theor. \textbf{43}, 075301 (2010). \end{thebibliography}
\end{document}